\documentclass[manuscript]{acmart}

\AtBeginDocument{%
  \providecommand\BibTeX{{%
    \normalfont B\kern-0.5em{\scshape i\kern-0.25em b}\kern-0.8em\TeX}}}

\newcommand{\etal}{\textit{et al}.}

\usepackage{amsthm}
\usepackage{amsmath}

\usepackage{thmtools}
\usepackage{thm-restate}
\usepackage{hyperref}

\allowdisplaybreaks
\newcommand{\name}{\textit{LangXRL}}

\usepackage{color}
\usepackage{mathtools}

\usepackage{subfigure}
\usepackage{multirow}
\usepackage{makecell}
\usepackage{array, booktabs, arydshln, xcolor}
\usepackage{graphbox}
\usepackage{tikz}
\usepackage{pgfplots}
\usepackage{wrapfig}
\usepackage{xspace}

\usepackage{amsmath,amsfonts,bm}

\def\eqref#1{equation~\ref{#1}}
\def\1{\bm{1}}

\DeclareMathAlphabet{\mathsfit}{\encodingdefault}{\sfdefault}{m}{sl}
\SetMathAlphabet{\mathsfit}{bold}{\encodingdefault}{\sfdefault}{bx}{n}

\begin{document}

\title{Decisions that Explain Themselves: A User-Centric Deep Reinforcement Learning Explanation System}

\author{Xiaoran Wu}
\affiliation{%
 \institution{Tsinghua University}
 \streetaddress{30 Shuangqing Rd.}
 \city{Beijing}
 \country{China}}

\author{Zihan Yan}
\affiliation{%
 \institution{Massachusetts Institute of Technology}
 \streetaddress{77 Massachusetts Ave}
 \city{Cambridge}
 \state{MA}
 \country{USA}}

\author{Chongjie Zhang}
\affiliation{%
 \institution{Tsinghua University}
 \streetaddress{30 Shuangqing Rd.}
 \city{Beijing}
 \country{China}}

\author{Tongshuang Wu}
\affiliation{%
 \institution{Carnegie Mellon University}
 \streetaddress{5000 Forbes Ave}
 \city{Pittsburgh}
 \state{PA}
 \country{USA}}

\renewcommand{\shortauthors}{anonymous \etal}

\begin{abstract}
With deep reinforcement learning (RL) systems like autonomous driving being wildly deployed but remaining largely opaque, developers frequently use explainable RL (XRL) tools to better understand and work with deep RL agents. However, previous XRL works employ a techno-centric research approach, ignoring how RL developers perceive the generated explanations. Through a pilot study, we identify major goals for RL practitioners to use XRL methods and four pitfalls that widen the gap between existing XRL methods and these goals. The pitfalls include inaccessible reasoning processes, inconsistent or unintelligible explanations, and explanations that cannot be generalized. To fill the discovered gap, we propose a counterfactual-inference-based explanation method that discovers the details of the reasoning process of RL agents and generates natural language explanations. Surrounding this method, we build an interactive XRL system where users can actively explore explanations and influential information. In a user study with 14 participants, we validated that developers identified 20.9\% more abnormal behaviors and limitations of RL agents with our system compared to the baseline method, and using our system helped end users improve their performance in actionability tests by 25.1\% in an auto-driving task and by 16.9\% in a StarCraft II micromanagement task.
\end{abstract}

% \sherry{and achieve more XX performance...What's the end goal of the user task?}
% \sherry{I think you would want to expand on the goal, otherwise it's a bit hard to decipher why e.g. temporal attention, LLM, NL are important in your solution}
% \sherry{doing....}
% \sherry{what's an experimenter? Participant? "excellent" is not a comparison over baseline. Maybe mention some numbers?}




\begin{teaserfigure}
    \includegraphics[width=\linewidth]{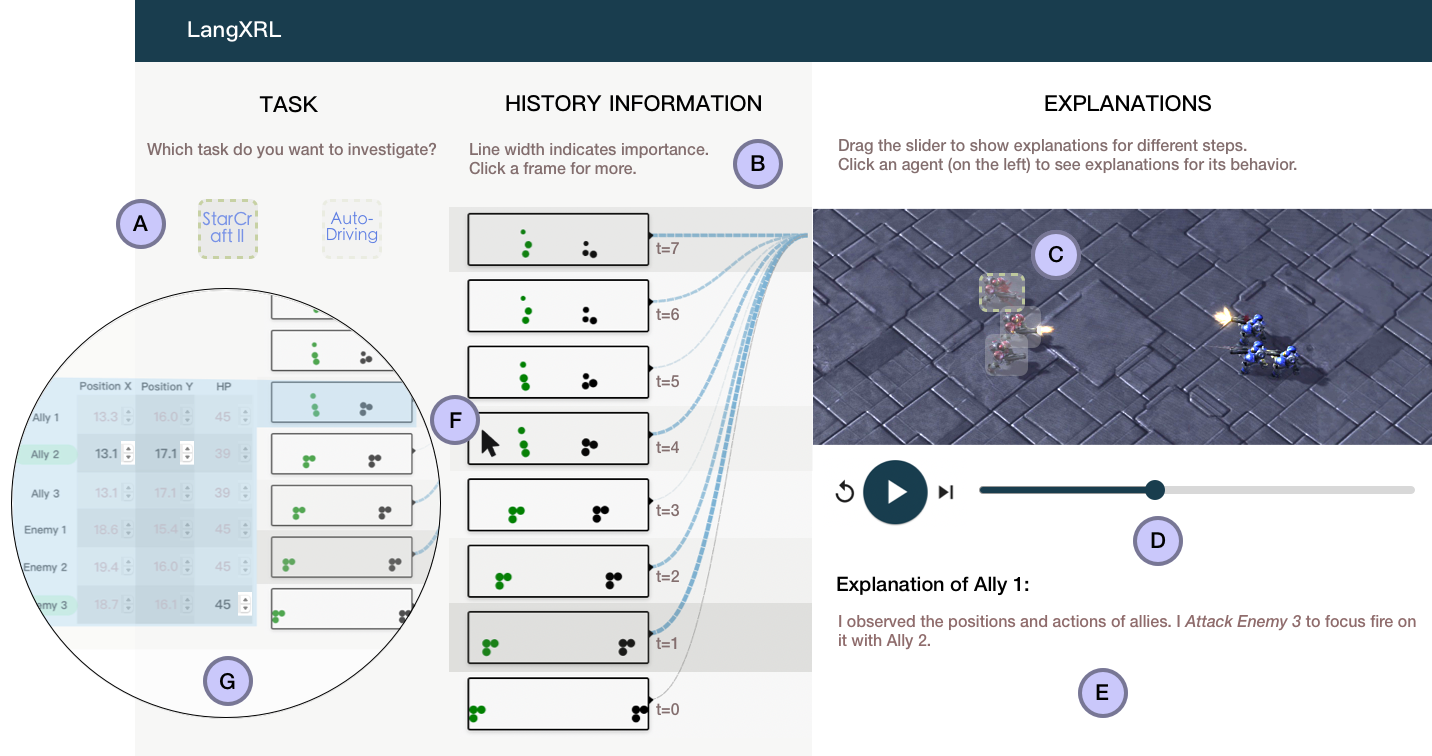}
    \caption{An overview of our interactive explainable reinforcement learning (XRL) system, \name. Users first choose a task (A), an agent (C), and a time step (D). Our system will use a novel XRL method to generate explanations in natural language (E), which are shown below the task scenarios (C, here the units on the left are controlled by RL, and the task is to defeat the units on the right). Important historical information that influence explanations is also shown (B), with line widths indicating importance. Users can dive in and see detailed influential information by clicking (F). They can also edit the history information and observe corresponding changes in agents' behavior, which allows active exploration of the reasoning behind RL decisions.}
    \label{fig:teaser}
\end{teaserfigure}

\maketitle

\section{Introduction}

Reinforcement learning (RL)~\cite{sutton2018reinforcement} is a machine learning paradigm that deals with sequential decision-making problems. An RL agent learns a policy to select actions that maximize expected cumulative rewards in the future. Using deep neural networks to represent policies~\cite{mnih2015human}, deep RL has made new forms of human-AI interaction possible in various human-in-the-loop decision-making scenarios, such as autonomous driving~\cite{wang2018deep}, robotic control~\cite{morales2021survey,zhu2021deep}, (stock) portfolio management~\cite{koratamaddi2021market}, and intelligent resource allocation~\cite{ning2021intelligent}.

Despite the progress, developing reliable deep RL models for human-AI interactions remains a major challenge~\cite{ibarz2021train,nguyen2020deep}. Unlike supervised learning where decisions only depend on current inputs, RL models are updated sequentially and accumulatively~\cite{franccois2018introduction}, i.e. they process information in the history of observation, and their decisions cause transition to new states ans new observation that will influence future decisions. As a result, the opaqueness caused by the black-box nature of neural networks is more severe in deep RL, making it more difficult for developers to correct abnormal decisions and find limitations for RL models.

% exacerbating users' unreliability of RL agents during the interaction.
% \sherry{ which is only trained onced on the labeled training dataset...explain in one sentence, don't assume IUI reviewers' knowledge level}
% With the development of explainable artificial intelligence (XAI), explainability has been proven to be a critical factor in
To alleviate the influence of opaqueness, developers use explainable RL (XRL) tools to better understand and work with deep RL agents. Previous works have proposed many XRL methods, like using simple interpretable models to mimic the behavior of deep RL agents~\cite{bastani2018verifiable,jhunjhunwala2019policy, bewley2021tripletree, liu2018toward} and using mathematical abstraction to represent the reasoning of RL models~\cite{hasanbeig2021deepsynth}. However, these works typically employ a techno-centric research approach, focusing more on the optimization of interpretable models. It is largely unclear how RL developers perceive these methods and whether these methods can indeed help build reliable RL agents that gain appropriate trust from end users.

% \sherry{The next sentence is directly jumping into the first contribution. Come up with one sentence that describe your overall contribution. e.g. We study the potential discrepancy between XRL design objective in \emph{research}, and end user expectations and needs in \emph{real use cases}.}
% \sherry{ah, so this is very developer focused, not really end user. Up until now some discrepancies in the writeup: 
% 1. You said end users will need to build trust in abstract, but then the identified goals are from developers' perspective. So I think you want to pick a user group.
% 2. You said you don't want people to misbehave when interacting with RL agents, which is different from ``enhancing trust''. You only want to enhance trust is the users are clearly undertrusing. Many prior work has shown trusting AI models too much could lead to overreliance and that's not a good thing - if the model makes mistake we wouldn't know to rewrite it.
% So at least we should say build appropriate trust.
% }

In this paper, we study and try to fill the potential discrepancy between design objective in XRL \emph{research}, and developer expectations and needs in \emph{real use cases}. We first carry out a formative study with 10 RL practitioners to understand their attitudes and interaction workflow with the existing XRL system. Through their self-reflections on own experience (and sometimes, frustration) interacting with XRL models, we identify three major goals for using XRL models: (1) \textbf{Abnormality detection}: To detect unexpected behavior of RL models under development and improve their reliability; (2) \textbf{Limitation assessment}: To understand the limitation of and thereby improve existing RL algorithms; and (3) \textbf{End-user support}: To deploy RL models with explanations so as to help their end users to understand and appropriately trust the model.

In addition, we summarize the following four pitfalls that prevent existing XRL methods from fulfilling these goals: (1) \emph{Parochialism}: users hardly get explanations for novel situations encountered in practice, hurting \emph{End-user support}. (2) \emph{Simplism}: important information like the reasoning process of the original model is inaccessible to the users, which hurts \emph{Abnormality detection}. (3) \emph{Subjectivism}: developers find inconsistency between explanations and RL models' capability, which hurts \emph{Limitation assessment}. (4) \emph{Formalism}: users face mathematical, formal, or abstract explanations~\cite{greydanus2018visualizing,atrey2019exploratory} that require high cognitive load to understand, hurting \emph{End-user support}. Moreover, the practitioners complain that previous XRL works put much stress on algorithm development but left a blank in terms of an user-friendly interface showing the correspondence between explanations and decisions.
% \sherry{Space wise I feel we spent much more time talking about formative studies but not much on system and user study. Maybe shorten the first several paragraphs, and expand the next one with a walkthrough case.}

Geared towards the above limitations of techno-centric XRL works, we present an intelligent XRL system underpinned by a novel XRL algorithm. We introduce a counterfactual inference approach to generate XRL explanations by identifying the causes and results of a decision -- see how the decision changes if the RL agents didn't observe an object and its action. This inference method benefits to avoid simplism and subjectivism because it reveals details of the reasoning process of the original RL agent. Furthermore, to avoid formalism, we build an influence graph that reflects interdependence between causes/results and the decision, which is translated into natural language. In our XRL system (\name, Fig.~\ref{fig:teaser}), the generated explanations are shown with the rendering of agents' behavior, and important information in observation history is highlighted to help users understand the dynamics of RL models. In addition, our system enables users to actively explore the information of interest by selecting RL agents, editing their observation, and then analyzing its behavior changes.

We carried out a user study with 14 participants, including 7 RL developers and 7 end users of their RL models, to evaluate whether \name~can help developers better achieve the three goals for using XRL identified in the formative study. All the participants used \name~to interact with RL agents. We validated that RL system developers identified $20.9\%$ more abnormal behaviors and limitations of RL agents with \name~compared to the baseline method. Moreover, their end users for the developed RL system reported that using \name~helps them distinguish the situations where RL agents do and do not work well, and thus, compared to the baseline, build a more appropriate trust with these agents. The participants then took part in explanation-actionability tests where they collaborated with RL agents to finish tasks. Using \name~improved the performance of such human-RL teams by $25.1\%$ in an auto-driving task and by $16.9\%$ in a StarCraft II\footnote{StarCraft II are trademarks of Blizzard Entertainment\textsuperscript{TM}.} micromanagement task. These results demonstrate that our XRL system helps the users know better about the behaviors of RL agents and build appropriate trust with them. 

In summary, our contributions are three folds:

\begin{itemize}
    \item A user-centric study that identifies the goals of RL developers for using XRL and four techno-centric pitfalls that prevent previous XRL methods from achieving these goals.
    \item An XRL algorithm based on counterfactual inference that detects causes and results of decisions, builds influence graphs reflecting the reasoning process of RL agents, and hereby generates explanations in natural language. 
    \item An interactive XRL system where users can actively explore the information of interests and the reasoning behind RL decisions.
\end{itemize}

\section{Related Works}
% \sherry{I think related work is usually either the second or the second last, can't appear between system and user study :)}

Explainable AI aims to produce insights about the cause of the models' decisions, make these decisions understandable by human users~\cite{arrieta2020explainable,carvalho2019machine,ehsan2019automated,gunning2017explainable,ras2018explanation}, and gain users' trust~\cite{gilpin2018explaining}. Most works on explainable AI study supervised learning, and only until recently has the research interests shifted to explainable RL. In the following sections, we discuss two major categories of previous work on explainable RL (XRL).

\subsection{Intrinsic Explainable RL}

Intrinsic XRL methods represent policies as inherently interpretable models. For example, \citet{silva2019optimization} use decision trees~\cite{rokach2005decision} as policy approximators. They make decision trees differentiable by replacing the Boolean decisions with sigmoid activation functions. \citet{topin2019generation} also adopt a decision tree policy, but they treat the action of choosing a feature to branch a tree node as additional actions in an augmented Markov Decision Process. A drawback of decision trees is that they can only produce axis-parallel partitions, whose weak representational capacity limits the interpretability of the model in real-world tasks. Additional mechanisms like algebraic expressions represented by tree nodes~\cite{hein2018interpretable, landajuela2021discovering} are developed to increase the capacity of decision tree policies. Nevertheless, these mechanisms inevitably increase the cognitive load from users to understand the generated explanations. The initial mental cost of understanding the explanation models~\cite{dodge2021no} makes these methods fall into \textsc{Formalism}.

\subsection{Post-hoc Explainable RL}
Post-hoc XRL methods try to relate the input (observations) and the output (actions) of a trained RL policy in an interpretable way. Some methods use an interpretable \emph{surrogate} model to approximate a learned policy model. Imitation learning~\cite{abbeel2004apprenticeship} and learning from demonstration~\cite{argall2009survey} are popular techniques for learning the surrogate model, which changes the RL problem into a classification problem -- the surrogate model learns to classify which action should be selected given the observation. Decision trees and their variations are the most common choice for the surrogate model~\cite{bastani2018verifiable,jhunjhunwala2019policy, bewley2021tripletree, liu2018toward}. Genetic programming~\cite{zhang2020interpretable} and programmatic policy searching~\cite{verma2018programmatically} are also adopted. Similarly, finite state machines~\cite{koul2018learning, danesh2021re} and deterministic finite automata~\cite{hasanbeig2021deepsynth} are used as surrogate models for recurrent neural networks. However, in order to be interpretable, surrogate models are designed as simple as possible to be understandable. Consequently, the representational capacity of these models typically cannot support them to interpretable all the decisions made by the original model. Therefore, these methods suffer from \textsc{Simplism}.

Some methods avoid using interpretable models with insufficient representational capacity by directly inserting explanation models into the original RL policy. Saliency maps distinguish the elements in the observation that influence the policy's decisions~\cite{atrey2019exploratory}. For example, \citet{greydanus2018visualizing} perturb the input by Gaussian blur and measure the corresponding changes in the policy output. \citet{gottesman2020interpretable} distinguish important experience that is influential to the learning process. However, these salience-based methods are viewed as insufficient for explaining RL. Salience only provides the focus of the model but does not capture the reasoning behind a decision~\cite{atrey2019exploratory}. This line of research relies on humans to give \emph{ad hoc} explanations after observing the saliency maps and thus falls into \textsc{Subjectivism}.

Methods use similar experiences encountered in the training process to explain the policy's decisions can alleviate subjectivism~\cite{lage2019exploring}. \citet{amir2018highlights} define state importance and diversity measurements to identify the most similar trajectory in the experience buffer. ~\citet{huang2018establishing} select similar trajectories where the chosen action has a much higher expected accumulative return. \citet{zahavy2016graying} learn a state embedding space where states that are close to each other are regarded as similar. Abstract policy graphs~\cite{topin2019generation} use expected future trajectories to summarize policies. These example-based methods highly rely on the experience collected during training time and typically depend on assumptions about the environment, such as binarized features~\cite{topin2019generation}. There is no guarantee about the interpretability of these methods when facing novel experiences or when the assumptions are relaxed. These drawbacks make them \textsc{Parochialism}.

Directly generating natural language explanations for the original RL model holds the promise of avoiding simplism, formalism, and parochialism. However, letting RL agents explain their actions in natural language is challenging because of the simultaneous learning of policy and language, and the alignment between them. \citet{ehsan2018rationalization} and~\citet{wang2019verbal} solve this problem by proposing a supervised learning framework using action-explanation pairs annotated by humans. However, since the explanations are provided by humans, these methods are actually learning how humans perceive the behavior of RL models and thus are \textsc{Subjectivism}. In this paper, we propose to generate natural language explanations by enabling the model to causally infer the causes of its behaviors and to express them in human languages, which avoids subjectivism.

In summary, previous XRL works mainly adopt a techno-centric research approach, and user-centric studies are largely absent from the XRL literature.

% Many real-world tasks feature partial observable environments, e.g., a self-driving car can only observe objects within the range of its sensors. To make decisions in such environments, RL models need to process  
% Some work pre-defines templates for the agent to fill in~\citet{hayes2017improving}. The former type typically enjoys more clarity but requires the explanation to adhere to a specified template. Furthermore, there is some question about the validity of these explanations.
% \begin{table}[]
%     \centering
%     \caption{Demographics of formative study participants.}
%     \vspace{-1em}
%     \begin{tabular}{c|cc|c}
%         \hline
%         ID & Gender & Age & Occupation \\
%         \hline
%         $P1$ & Male   & 37 & RL algorithm designer for a recommender system \\
%         $P2$ & Female & 29 & RL algorithm designer for a recommender system \\
%         $P3$ & Female & 26 & Game AI engineer\\
%         $P4$ & Male   & 41 & RL researcher (professor) \\
%         $P5$ & Female & 26 & RL researcher (graduate student) \\
%         $P6$ & Male   & 28 & Engineer in an intelligent wind farm \\
%         $P7$ & Female & 23 & Engineer in an intelligent wind farm \\
%         $P8$ & Female & 24 & Teaching assistant of a deep RL course \\
%         $P9$ & Male.  & 29 & Auto-driving engineer \\
%         $P10$ & Female & 43 & Auto-driving decision model developer \\
%         \hline
%     \end{tabular}
%     \label{tab:formative_demograpgics}
% \end{table}

\section{Formative Study}\label{sec:fs}

As aforementioned, although many previous works focus on enhancing the explainability of reinforcement learning (XRL), a human-centric study is still not fully explored. It is largely unclear what RL practitioners are expecting from XRL methods and whether these methods can fulfill their needs. Therefore, in this section, we engaged in a formative investigation to identify the insufficiency of existing XRL methods in common usage scenarios with specific using goals and inform design opportunities that address these challenges.

\subsection{Method}
% \sherry{The following participant paragraph is a bit of a laundry list right now. If this is important, say you tried to cover different use cases (though most of them still seem like designers) and different applications. For example you can say "These practitioners work on X different domains including X, X, X, etc." If not very important just simplify}
We interviewed $10$ RL practitioners who frequently use XRL methods in their work (4 female and 6 male with an average age of 31). These practitioners work on 5 different domains including recommender systems powered by RL algorithms, RL-based game AI designing, RL research, RL-controlled intelligent wind farm, and autonomous driving.

% $P1$ and $P2$ are algorithm designers for a . $P3$ is a game AI designer and her AI is trained by RL algorithms. $P4$ and $P5$ are RL researchers, developing novel RL algorithms and models. $P6$ and $P7$ are engineers in an intelligent wind farm. An RL algorithm controls the orientation of wind turbines in the farm, and $P6$ and $P7$'s work is to detect and correct sub-optimal actions of the RL agents. $P8$ is a teaching assistant of deep reinforcement learning course. He uses XRL models to help students understand abstract concepts like Q-function and policy gradients. $P9$ and $P10$ are engineers of autonomous driving. The auto-driving vehicle is controlled by an RL agent, $P6$ and $P7$ use XRL models to probe the behavior of the auto-driving vehicle. 

Our interview session adopted a semi-structured format, in which we asked our participants about (1) what their goal of using XRL models is; why would a developer want to interact with an RL system with explanations?  (2) what XRL models they frequently use; (3) what is their general attitude towards existing XRL methods; and (4) do they have suggestions for improving the XRL models. The interviews lasted 30 minutes for each participant and were carried out via video conferencing. We placed fewer restrictions on the interviewees' responses to gain as much information as possible. We applied thematic analysis~\cite{braun2012thematic} to analyze the interview transcripts of the recorded videos. Two researchers coded these data individually and then discuss to converge findings.

% Do they want high accuracy on some decision making task? Do they want to correct their models?
% Need to add analysis method, for example:

\subsection{Insight 1: Goals for using XRL methods}

We first summarize three major goals of RL practitioners to use XRL tools.

\emph{\textbf{Goal 1: Abnormality detection.} Detect and correct problematic RL decisions and improve the decision models' accuracy.} RL models are expected to learn control policies that is more effective than humans in some complex applications like game AI and recommender system design. Human engineers should find out flaws of RL models so that corresponding samples can be collected to improve decision accuracy. To achieve this goal, human engineers typically "\emph{observe the behavior of the RL agent and select the decisions that seem irrational to humans}" ($P7$). However, for these decisions that are not intelligible to humans, it is difficult to "\emph{distinguish whether it is because the RL model makes a bad decision or is because the decision is correct but sophisticated}" ($P3$). By checking the explanations provided by the RL model, "\emph{engineers can check whether there is a persuasive reasoning behind each decision and thus better detect wrong decisions}" ($P1$).

\emph{\textbf{Goal 2: Limitation assessment.} Understand the limitation of current RL models and thereby develop better RL algorithms.} For example, $P10$ said: "In our project, the auto-driving vehicle frequently collides with other cars in roundabouts. After checking the explanations, we find that it believes it is at an intersection but can't find traffic lights. We thus design a special sub-module for navigating the roundabouts." Another example is $P5$: "I added a salient map module to my RL model for StarCraft II micromanagement tasks. Then I found that the model focused on useless background information, which may be because of a misleading intrinsic reward setting. So I changed my algorithm, and it worked."

\emph{\textbf{Goal 3: End-user support.} Build appropriate trust between end users and RL decision system by helping users understand.} In some real-world applications, even a tiny proportion of sub-optimal decisions can cause serious consequences. In these applications, explanations are critical in building an appropriate trust. On the one hand, end users feel reliable when "\emph{they see and agree with the explanations of RL models. For example, an auto-driving vehicle keeps its speed when there is a pedestrian crossing the road. The human driver needs to know whether, in the vehicle's calculation, a collision is impossible given the current movement status or the RL model simply ignores the pedestrian}" ($P9$). On the other hand, explanations can build an appropriate trust because they make the RL agent controllable: "\emph{human can override an RL decision without reasonable explanations to avoid a possible costly consequence}" ($P6$).
% \sherry{
% 1. Try to rephrase the intro part more like this. When you say "to deploy together and enhance user trust" it sounds too much like companies hoping their customers will happily rely on their auto-driving system.
% 2. Reorder goal 2 and 3? So 1-2 are developer and 3 is end user?
% 3. You should move the discussion in user study to here, and say "if look closer there are two targeted user group here." And did they have insights on how the need would be different based on the goals? I think it's a bit abrumpt to just end on Goal 3 and move to next section, usually there's some sort of discussion}

\begin{table}[]
    \centering
    \caption{Techno-centric pitfalls found in our formative study.}
    \vspace{-1em}
    \begin{tabular}{p{0.1\linewidth}p{0.42\linewidth}p{0.4\linewidth}}
        \toprule
        Pitfall & Manifestation & Cause \\
        \toprule
        Simplism & Summarize behaviors without showing underlying reasoning and details & Use a simple but interpretable model to mimic the behaviors of RL models \\
        \hline
        Parochialism & Only works for some cases and cannot generalize to novel situations & Limited model capacity or training data \\
        \hline
        Subjectivism & Inconsistent explanations and behavior & Train by action-explanation pairs provided by human observers\\
        \hline
        Formalism & Require significant cognitive load. Difficult for the users to understand  & Generating explanations in abstract or mathematical format \\
        \bottomrule
    \end{tabular}
    \label{tab:formative_demograpgics}
\end{table}

% Parochialism occurs when an XRL model cannot generalize explanations beyond a certain situations (e.g.,~\cite{verma2018programmatically,topin2019generation})
% Simplism is caused by using simple models with reduced representational capacity to mimic the deep policies~\cite{bastani2018verifiable,bewley2021tripletree,koul2018learning}. 
% Subjectivism happens when action-explanation pairs provided by human observers are used to train the explaining model~\cite{ehsan2018rationalization} and human understandings override the true dynamics of the model
% In this case, users find it hard to build appropriate trust to achieve \emph{goal 2}. 

\subsection{Insight 2: Pitfalls of existing XRL methods from a human-centric perspective}\label{sec:pitfalls}

Participants mentioned a lot of dissatisfaction when talking about their general altitude and suggestions for existing XRL models, which we summarized into four pitfalls.

\emph{\textbf{Pitfall 1}: Simplism}. In some XRL methods, explanations only summarize RL agents' behavior, and some valuable original information and the reasoning process of RL agents are inaccessible. Simplism happens when using simple models with reduced representational capacity to mimic the deep policies~\cite{bastani2018verifiable,bewley2021tripletree,koul2018learning}. Some participants ($P2$, $P7$, and $P10$) reported that their XRL models generate explanations that are superficial. For example, $P7$ said: "\emph{The explanations are just repeating what the agent has done in last few steps, without any comments on the reasons.}" $P2$ found the XRL model can only explain some simple behaviors with obvious reasoning: \emph{In games, the model often says: "Unit A attacks because it sees Unit B." Such explanations are useless. By contrast, in some cases where the units' actions are sophisticated, the model provides an obviously incorrect explanation.} In $P2$'s case, the policy represented by neural networks is distilled into an interpretable model of decision trees~\cite{bastani2018verifiable}. Other simple XRL models include finite-state machines~\cite{koul2018learning} ($P10$). These surrogate models have limited representational capacity and cannot fully understand the original policy.

Simplism hurts \emph{Abnormality detection} because such XRL models barely explain reasoning between decisions. Thus, humans can hardly determine whether an RL decision is reasonable or not. It also hurts \emph{End-user support}, which is obvious from the participants' comments -- such XRL models do not provide many insights that help humans understand the RL model, and trust cannot be built when users only obtain limited information. 

\begin{figure}
    \includegraphics[width=\linewidth]{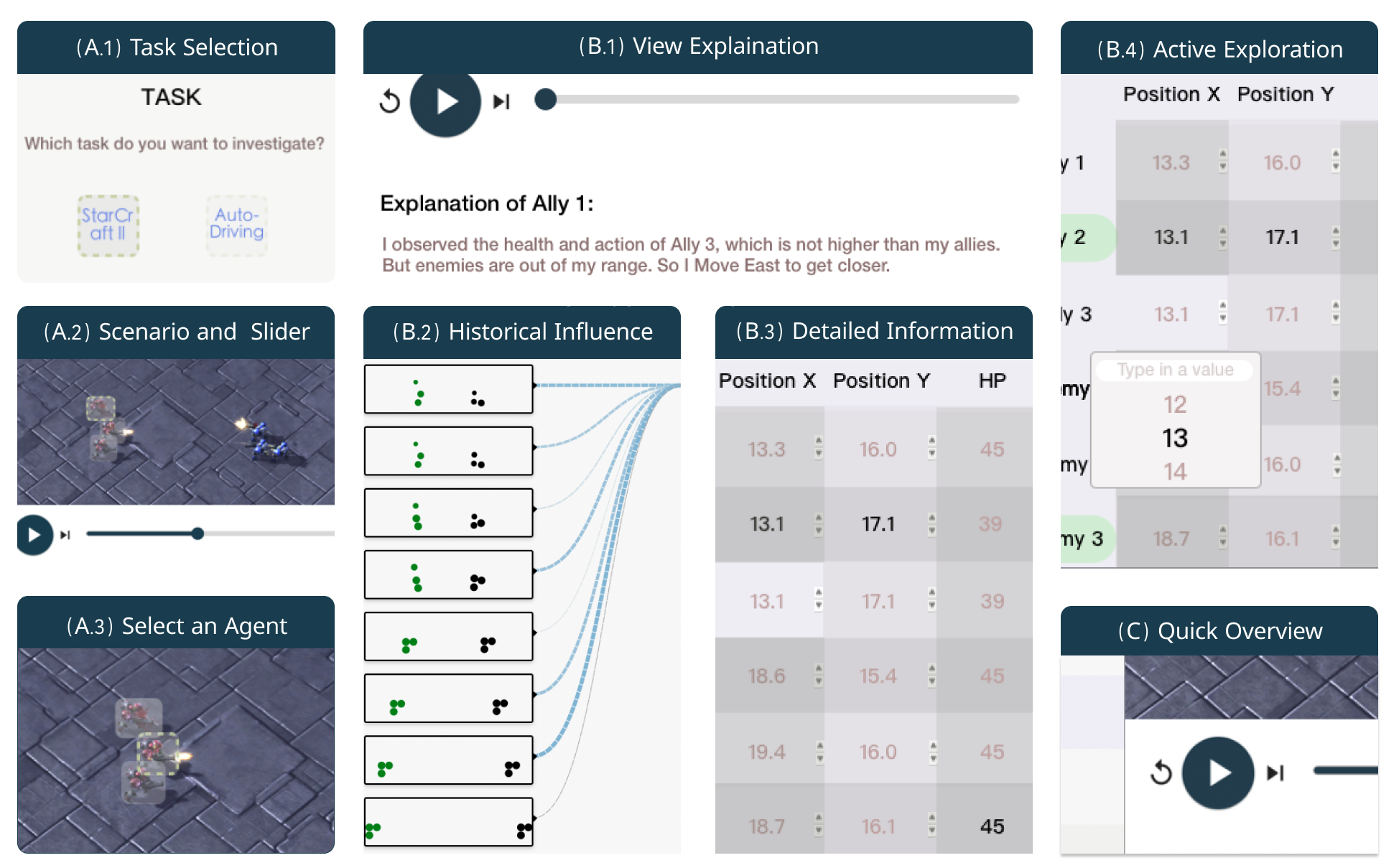}
    \caption{User interactions provided by \name. Users can first select the task that they want to study (A.1). Then, by dragging the slider (A.2) and clicking in the corresponding scenarios (A.3), users select the target time step and the target agent. Explanations for the target agent and time step will appear (B.1). These explanations are generated by our novel XRL method introduced in Sec.~\ref{sec:algorithm}. There are a lot more historical details for helping people to dive in (B.2), where the line widths indicate the importance of information in historical time steps. Users can further dive in to see the details (B.3) and modify them to see the change in agents' behavior (B.4). If users want to get less information and only see explanations in important time steps, they can click the \emph{play} button (C).}
    \label{fig:system}
\end{figure}

\emph{\textbf{Pitfall 2}: Parochialism}. It is difficult to gain explanations for novel situations. Participants ($P1$ and $P4$) found that their XRL models work satisfactorily in some cases, but not in others. For example, $P4$ said that: "\emph{I used an XRL model for a DQN~\cite{mnih2015human} agent playing Atari games. It works fine at the first level of Montezuma's Revenge, but not in all the games.}" $P1$ reported that his RL recommender system "\emph{works for some items, but provides wrong explanations for others}". We investigated the XRL methods used by $P1$ and $P4$, and found that parochialism is caused by the lack of training data or some unrealistic assumptions about tasks. For example, $P4$'s model works well when state features can be binarized~\cite{topin2019generation}, and $P1$'s model was trained on a few state-action pairs~\cite{verma2018programmatically}. Users can hardly get explanations for novel situations in real-world problems. 

Parochialism significantly prevents \emph{Abnormality detection} because in cases where the XRL model can not provide correct explanations, human users are not able to detect possible problems of the RL policy. Parochialism also hurts \emph{End-user support} because even failures in a small proportion of situations would make humans feel uncertain about the model in other situations, as $P1$ said: "\emph{The XRL model fails in one roundabout. Although I am not sure whether the RL algorithm works well at other situations, my users tend to be skeptical about its correctness. Being careful is always good in designing an AI model.}"

\emph{\textbf{Pitfall 3}: Subjectivism.} Some XRL systems provide explanations that are inconsistent or beyond the capability of the RL models. Participants $P3$, $P6$, and $P9$ reported a discrepancy between explanations and the RL model's capability. For example, $P9$ said: "\emph{The XRL algorithm said that the auto-driving vehicle accelerated because the next traffic light was about to turn green, and the auto-driver was concerned that the light would stay green for a short of time. I don't believe our RL policy can do such sophisticated inference, and the true reason might be simple -- it was a downhill road.}" We checked the explanation generation modules used by $P3$, $P6$, and $P9$, and found that they are trained by action-explanation pairs provided by human observers as groundtruth~\cite{ehsan2018rationalization}. In such cases, human understanding could override the true decision-making dynamics of the RL model. 

Subjectivism prevents \emph{Limitation assessment} because it sometimes leads to an overestimation of the RL model's ability, making it difficult to pinpoint its drawbacks. For example, $P6$ said: "\emph{My XRL model seems to know more than my RL model. In one situation, the XRL model gives an explanation for a decision. However, in a similar situation, when the XRL gives a similar explanation, the RL model actually makes an opposite decision.} Subjectivism also hurts \emph{End-user support} because the misalignment between explanations and RL decisions makes human users feel uncertain.

\emph{\textbf{Pitfall 4}: Formalism.} Some XRL methods provides mathematical, abstract, or indirect RL model explanations that require high cognitive load. Some participants ($P5$ and $P8$) are dissatisfied with their XRL tools because the generated explanations are abstract and difficult to understand. $P8$ said: "\emph{My explanation model generates saliency maps highlighting influential elements in the observation. Every time I presented the maps to my end users, I had to elaborate on how the maps worked. Even if the users know saliency maps, it is not obvious how the highlighted elements influence the final decisions.}" Formalism occurs when the XRL model seeks indirect or abstract mathematical solutions, like saliency maps~\cite{greydanus2018visualizing,atrey2019exploratory} or deterministic finite automata~\cite{hasanbeig2021deepsynth}. 

Formalism hurts \emph{End-user support} because end users may be unable to understand and absorb the provided information within a limited time budget. As stated by $P5$: "\emph{When the user does not have the patience or corresponding background knowledge, it gets tends to believe and overly rely on the XRL model. Such a trust is not appropriate and may be dangerous.}" 

\subsection{Insight 3: Design Strategies}\label{sec:ds}

Inspired by the goals for using XRL methods and the pitfalls faced by previous XRL methods, we identify four design strategies that will guide the design of our XRL system and algorithms.

\emph{\textbf{Design strategy 1}: Make the XRL model truthfully reflect the reasoning process of the RL algorithm.} This strategy can help reduce the possibility of simplism and subjectivism. 

\emph{\textbf{Design strategy 2}: Present explanations in an easily understandable format, such as natural language.} This strategy can avoid abstract explanations and thus prevent formalism. 

\emph{\textbf{Design strategy 3}: Adjust the explanation generation model to suit all scenarios and be applicable to most RL algorithms,} which can effectively increase the generalizability of XRL and thus reduce parochialism.

\emph{\textbf{Design strategy 4}: Present explanations in an interactive manner, enabling users to actively choose what they want to learn about the RL model.} This design strategy is to better fulfill the three goals for using XRL models. Users can study cases of interest to accurately pinpoint possible problems of the underlying reinforcement learning algorithm, which could help build an appropriate trust and find the limitation of the RL model.
\section{\name}

% Motivated by the findings and design strategies inspired by the formative study, we now introduce our XRL system (Sec.~\ref{sec:system}) and the underpinning novel explainable RL algorithms (Sec.~\ref{sec:algorithm})

Following the aforementioned four design strategies, we designed and implemented \name, an AI-enabled interactive tool for studying and understanding the underlying dynamics and reasoning of RL algorithms. \name~presents features that allow active exploration of RL agents' behavior. In Sec.~\ref{sec:ui}, we introduce these features, and in Sec.~\ref{sec:algorithm}, we describe our XRL algorithm that underpins these features.

\subsection{System User Interface}\label{sec:ui}

% \sherry{
% 1. I think it might be better to have an actual example in figure 1 where people can get a concrete idea of what interactions they would perform and why would they want to do that. Right now it's a bit vauge.
% 2. As a system paper there's usually some interface novelty + some algorithm novelty. The way it's currently written feels like interface is quite basic and you dont care much about it :) But it's actually a pretty complicated interface, so I'd put more effort to explain:
% 1. senario
% 2. below direct simple explanation so it's...the expl is generated by al aogorithm below
% 3. There are a lot more details for helping people to dive in on the left
% - We present F so people have context of historical info which is important because...
% - we do detail on demand for G (which is what?)

% }
The features of \name~are summarized in Fig.~\ref{fig:system}. After selecting a task in Part A, our system shows and explains RL agents' behavior in an episode of the task. The agents face a sequential decision-making problem that requires it to choose an action in each time step. The replay of these decisions, as well as the corresponding task scenarios, are shown in the upper right part of the interface. 

By dragging the slider (Part D), users can select time steps. For each time step, an explanation in natural language will appear in Part E, interpreting the agent's decision at this time step. These explanations are generated by a novel method that avoids the four user-centric pitfalls introduced in Sec.~\ref{sec:pitfalls}. We will describe our explanation-generation method in detail in Sec.~\ref{sec:algorithm}. When there is more than one RL agent, users can choose a \emph{target agent} by clicking it in Part C. Other parts of the interface will change to explain the behavior of the target agent.

An RL agent's behavior typically depends on its observation history, e.g., an auto-driving vehicle needs camera input data in the last few steps to decide a pedestrian's moving direction. Correspondingly, our explanations are based on the analysis of the agent's observation history. To help users better explore and understand the agent's behavior and our explanations, in Part B, we provide sketches for observation in recent time steps where influential historical time steps are highlighted by deep background color and a wide curve.

If users are curious about which information in a recent time step leads to the final decision of the RL agent, they can click the corresponding rectangle in Part B. An information list would appear, where important information is shown in bold font. Users can also interactively change the information here in the information list (Part G), and our system will calculate and render agents' behavior under the new configuration. These interactions are underpinned by a novel counterfactual inference method that quantifies the importance of information, which will be discussed in Sec.~\ref{sec:graph}.

If users do not need much detail and just want to have a quick glance of an agent's behavior with explanations, they can click the \emph{play} button. The game states for each time step (Part C) and explanations (Part E) for important decisions will be played. In this way, users can see important information in an online fashion within a limited time budget.

\begin{figure}
    \includegraphics[width=\linewidth]{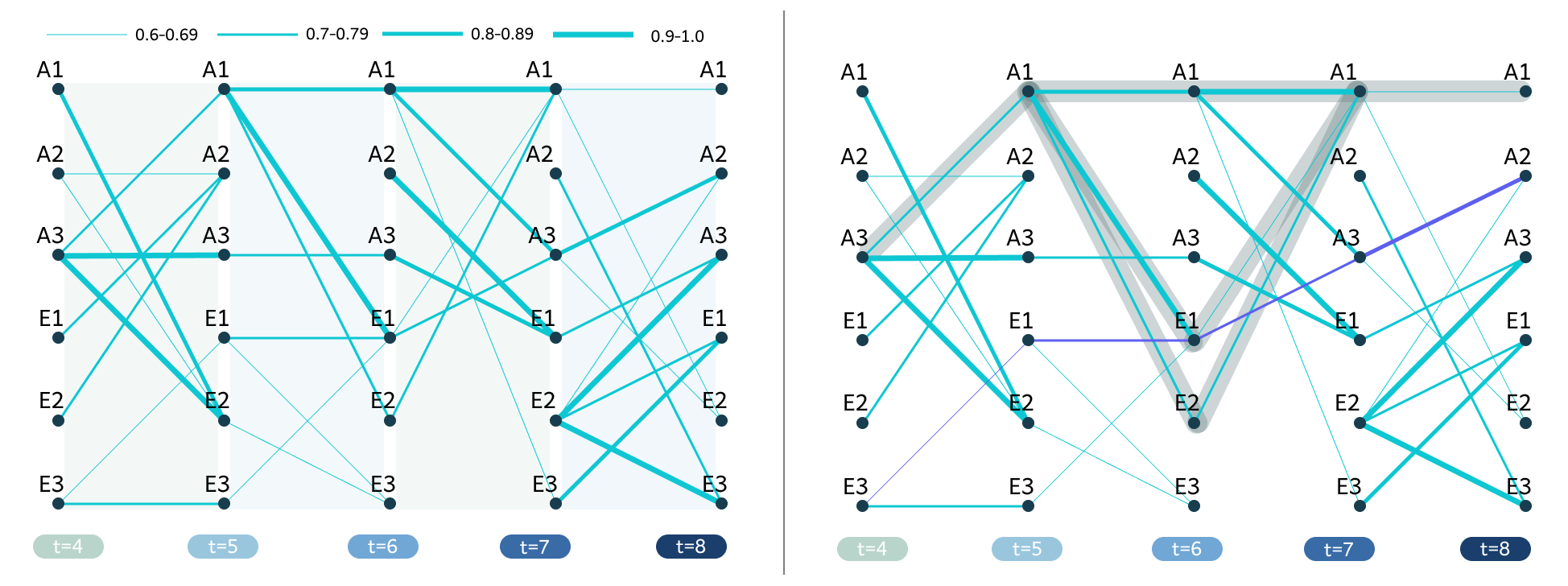}
    \caption{Left: An example of influence graphs built for generating explanations on StarCraft II micromanagement task. A1 stands for Ally 1, and E1 stands for Enemy 1. The weights of edges are the influential values calculated using Eq.~\ref{equ:influence}, and larger values indicate stronger influence. Right: A maximum flow (highlighted by the gray color) for calculating the influence of Ally 3 at time step 4 on Ally 1's decision at time step 8. Also shown here is a longest path (highlighted by the purple color) for identifying the causes of the decision of Ally 2 at time step 8.}
    \label{fig:method}
\end{figure}

\subsection{Computational Methods}\label{sec:algorithm}
The interactions introduced in the previous section are supported by a novel explainable RL method. In this section, we describe the technical details and novelties of this computational method.

For a quick overview, our explanations are generated by finding the decision's reason and result. The reason and the result is found by counterfactual inference (Sec.~\ref{sec:counterfactual}): what would happen if an object did \emph{not} do an action? The advantage of such inference is that it does not depend on specific tasks and RL algorithms and is applicable to all kinds of RL learning settings (Design strategy 2). The influence between objects and actions are organized by constructing an influence graph (Sec.~\ref{sec:graph}) reflecting the reasoning process of the original RL model (Design strategy 1), from which natural language explanations (Sec.~\ref{sec:language}) can be easily extracted (Design strategy 3).

% \sherry{we also need an overview on the 4.2.1-4.2.3 so people know what the algorithm is doing at a high level, e.g. what's the input and output per step}
% \sherry{Similar to background, this gets to the tech detail a bit too deeply, We really need to add more intuitions on why these would work.}

\subsubsection{Counterfactual influence analysis}\label{sec:counterfactual}

We first discuss how to quantify the influence of an object and its action on a given decision, which can help us find the reason and cause of the decision. 

A prerequisite of such influence quantification is to detect objects. Object detection is extensively studied in the field of computer vision. Mask R-CNN~\cite{he2017mask} is a classic and accurate deep model that predicts the object labels and provides pixel-level masks showing the detailed position of objects (instance segmentation). Formally, given an observation $o$ (which is an image) of the size $h\times w$, we use the instance segmentation head of a trained Mask R-CNN model that outputs $k$ masks $\{m^1, m^2,\dots,m^k\}$ for $k$ detected elements. Each mask $m^i\in\mathbb{R}^{h\times w}$ has the same size as the input image. Masks are binary, with $1$ meaning that the corresponding pixel belongs to the element and $0$ meaning the opposite.

With detected objects, we now discuss how to quantify their influence on a decision. Suppose that, at time step $t$, under the original observation history $\tau_t\equiv(o_1, o_2, \cdots, o_t)$, the RL model makes a decision that is represented as a distribution over the action space $\pi(\cdot | \tau_t)$. To model the influence of an object $i$ and its action at time step $t$, we replace the observation $o_t$ with sub-images where object $i$ is masked: $o_t^{-i}=o_t \circ (1-m^i)$, where $\circ$ is the matrix element-wise multiplication. Replacing $o_t$ with $o_t^{-i}$, we get the observation history $\tau_t^{-i}$ where information about object $i$ is modified. Then the influence of object $i$ and its action on the current decision can be calculated as: 
\begin{equation}
    I^i_t = D_{\mathtt{JS}}\left[\pi(\cdot | \tau_t) \| \pi(\cdot | \tau^{-i}_t)\right],\label{equ:influence}
\end{equation}
where $D_{\mathtt{JS}}$ is the Jensen-Shannon divergence between two distributions. An object with a higher $I^i_t$ value exerts more significant influence and is more likely to be a reason of the current decision. Similarly, if we mask the target agent and observe changes in other objects' behavior in the next time step, we can identify the cause of the current decision. 

\subsubsection{Construct influence graphs}\label{sec:graph}
As discussed before, RL decisions are largely dependent on historical observation information. Intuitively, we can extend the technique described in Sec.~\ref{sec:counterfactual} to include history analysis by masking object $i$ in all previous time steps. However, this method only considers the direct influence of object $i$ and ignores the indirect influence exerted through other objects in the environment. Another drawback of this method is computational complexity, which is $O(kT^2)$ where $k$ is the number of objects and $T$ is the total time steps in an episode of the task. This is because, for each time step, we need to calculate the influence of each object from the beginning of the episode. 

Constructing an influence graph can consider both direct and indirect influence while reducing the time complexity of influence analysis to $O(kT)$. An influence graph (Fig.~\ref{fig:method}-left) consists of multiple layers where each layer represents a time step. We start from the beginning of the episode. At each time step, we add a new layer to the influence graph. Each node in this new layer represents an object detected in the observation. Then we use Eq.~\ref{equ:influence} to calculate the influence of objects in the previous layer on the objects in the current layer. An edge is added between these two layers connecting two objects whose $I_t$ value is larger than a threshold $\xi$. The weight of this new edge $I_t$.

We now discuss how to calculate the influence of an object on the current decision within a constructed influence graph. Given an object $i$ at time step $t_1$ and a decision at time step $t_2$, we calculate \emph{the maximum flow}~\cite{goldberg1988new} (Fig.~\ref{fig:method}-right) from the source node (node $i$ at the layer $t_1$) to the sink node (the RL agent node at the layer $t_2$). The maximum flow problem is among the most classic problems in algorithm design. Intuitively, the value of the maximum flow here quantifies the maximum possible influence exerted through interplay among all objects in the task, which aligns perfectly with our aim. In practice, these values are used to highlight important time steps and information in Part B, F, and G of our system (Fig.~\ref{fig:system}).

\subsubsection{Generate Natural Language Explanations}\label{sec:language}
Another advantage of building influence graphs is to provide a structure from which explanations can be extracted. Intuitively, an explanation mainly answers a \emph{why} question -- why does the agent make this decision. Such a question can be answered by providing the reason and result of the decision, which is encoded in the influence graph.

Specifically, for a given decision, we use the technique introduced in Sec.~\ref{sec:graph} to identify object $i$ that influences the decision most significantly, and object $j$ that is influenced most significantly by the decision. We then find the paths with maximum accumulated weights connecting from $i$ to the RL agent (Fig.~\ref{fig:method}-right), and from the RL agent to $j$. Then the explanations are generated by a template: \emph{I observed [object $i$' and its behaviors or properties], so I [the current decision] to [object $j$'s change after the current decision].} Object $i$ and $j$'s behavior or change can be decided by tracking the paths. For example, in autonomous driving, by tracking the path, we may find the pedestrian's $x$-coordinates is increasing and we know it is moving rightward. The final explanation can be \emph{"I observed [a pedestrian is moving rightward], so I [brake] to [avoid the pedestrian.]"}

% \sherry{Do you have any auto-eval you can paste here?}
\begin{figure}
    \centering
    \includegraphics[width=\linewidth]{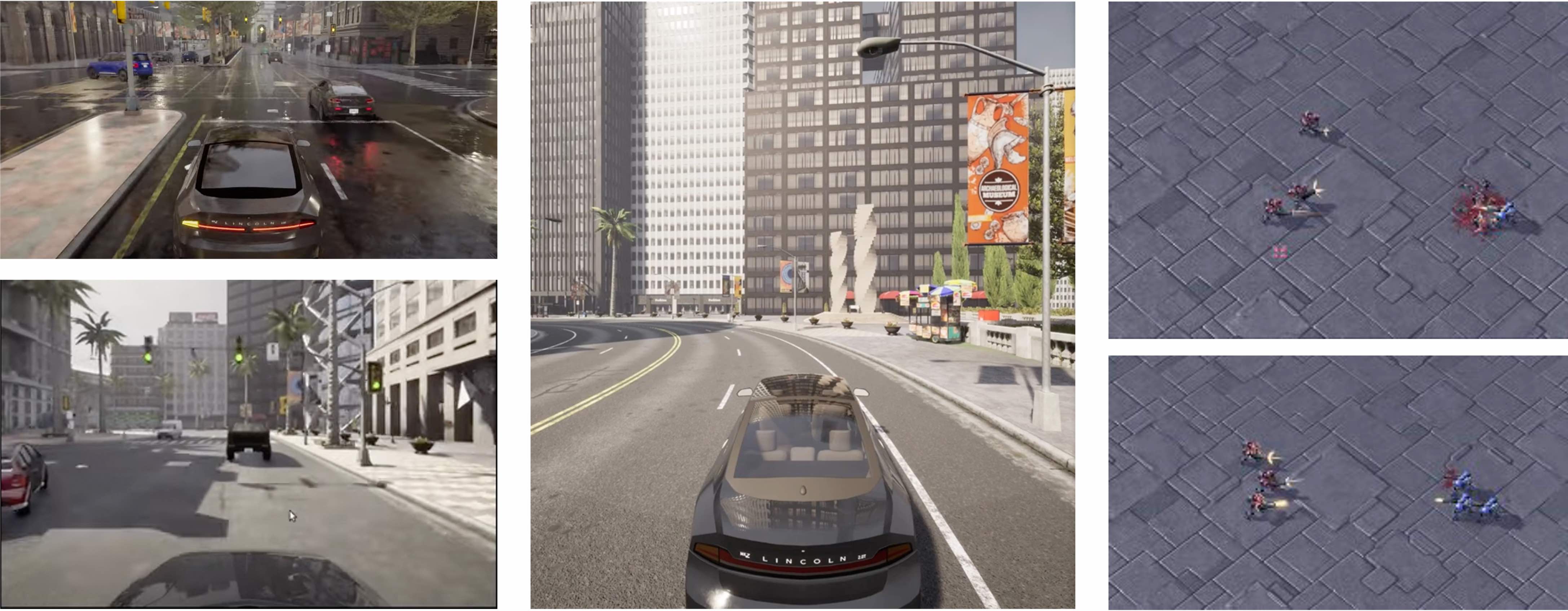}
    \caption{Demonstrative figures for the two tasks tested in our user study. Left: Auto-driving; Right: StarCraft II micromanagement. In the auto-driving task, RL agents control vehicles and are tasked to reach destinations as soon as possible without collisions. In the StarCraft II task, RL controls allies on the left to defeat the enemies on the right.}
    \label{fig:snapshot}
\end{figure}

\section{User Study}

We conducted a user study to evaluate \name. The study mainly evaluated to what extent \name~can help developers achieve their goals of using XRL and avoid four techno-centric pitfalls summarized in the formative study (Sec.~\ref{sec:fs}). Specifically, the user study examined the following research questions:

\begin{itemize}
    \item \textbf{RQ1}: To what extent \name~helps developers identify abnormal behaviors (\emph{Goal 1}), find limitations (\emph{Goal 2}), and build appropriate trust between end users and RL models(\emph{Goal 3})?
    \item \textbf{RQ2}: Do users find explanations generated by \name~helps avoid the pitfalls of the techno-centric XRL methods?
    \item \textbf{RQ3}: Do users find the interactions provided by \name~useful and likable?
\end{itemize}

\subsection{Participants}

We recruited 14 participants (8 male, 6 female) aged 21 to 47 years old ($\mu$=29). Among them, 7 participants (3 female and 4 male with an average age of 32) developers of RL models. 2 of them are graduate students, 1 is a professor, and 4 are engineers working with RL. The other 7 participants (3 female and 4 male with an average age of 26) are end users of the RL models developed by these developers. Each participant was compensated with \$15 gift card for their time. Our user study protocol was approved by the IRB at our institution.

\subsection{Tasks \& Baseline}

For each participant, the study was carried out on two tasks (demonstrative figures are shown in Fig.~\ref{fig:snapshot}). The reason why we use two tasks is to avoid user biases -- some users may be familiar with one task which will influence the outcomes of their interaction with XRL methods.

\begin{itemize}
    \item Task 1: an auto-driving task. The simulator is built on Carla~\cite{Dosovitskiy17}, and we use the Town10-HD map. We generated 150 auto-driving vehicles and 30 pedestrians on the map. The auto-driving vehicles are controlled by RL agents whose policies are trained to reach destinations as soon as possible while avoiding collisions. 
    \item Task 2: StarCraft II micromanagement. In this task, RL agents control a team of units to fight against an enemy team controlled by built-in AI provided by the game engine. The RL agents are trained to win the game by dealing as much damage as possible to the enemy units. The simulator is based on PyMARL~\cite{samvelyan19smac}, and we used the map $\mathtt{3m}$.
\end{itemize}
% \sherry{Need demonstrative figures for the two use cases}

% Since the goal of using XRL models varies with the background of participants, the study setups for the two groups of participants were different. For the first group, the focus was on whether the XRL method or system can help participants find the abnormal behaviors (\emph{Goal 1}) and limitations (\emph{Goal 2})\sherry{this was your goal 3 xD} of the underlying RL models. Therefore, we selected 10 time steps in the task and intentionally overrode the decisions of the RL agents at these steps. Participants were tasked to report the abnormal behaviors of RL agents after using the baseline and our system.
% \sherry{Did you explain hte other group or did I miss it?}

We compared our system with a baseline method~\cite{ehsan2018rationalization}. As aforementioned in the related work section, this method also generates explanations in natural language but uses human-provided explanations for supervised learning. Since no XRL tools accompany this method, we presented action-explanation pairs in plain text format to the participants during testing. After installation, each participant used the baseline method and \name~to see explanations for the behavior of an RL agent. 

% \subsection{Data \& Apparatus}

% To test the generalizability of the explanation methods, the RL agents are trained with different RL algorithms. We considered PPO~\cite{schulman2017proximal}, deep Q-learning~\cite{mnih2015human}, and SAC~\cite{haarnoja2018soft}. The match between RL agents and participants was random.

\subsection{Procedure}

For each participant, the whole study session lasted for around an hour. And we list the detailed procedure below.

\subsubsection{Pre-task.}

Firstly, we prepared a tutorial on how to install and use \name. Before the beginning of each session, participants read the tutorial and installed \name~on their computers under the authors' guidance.

\begin{figure}
    \centering
    \includegraphics[width=\linewidth]{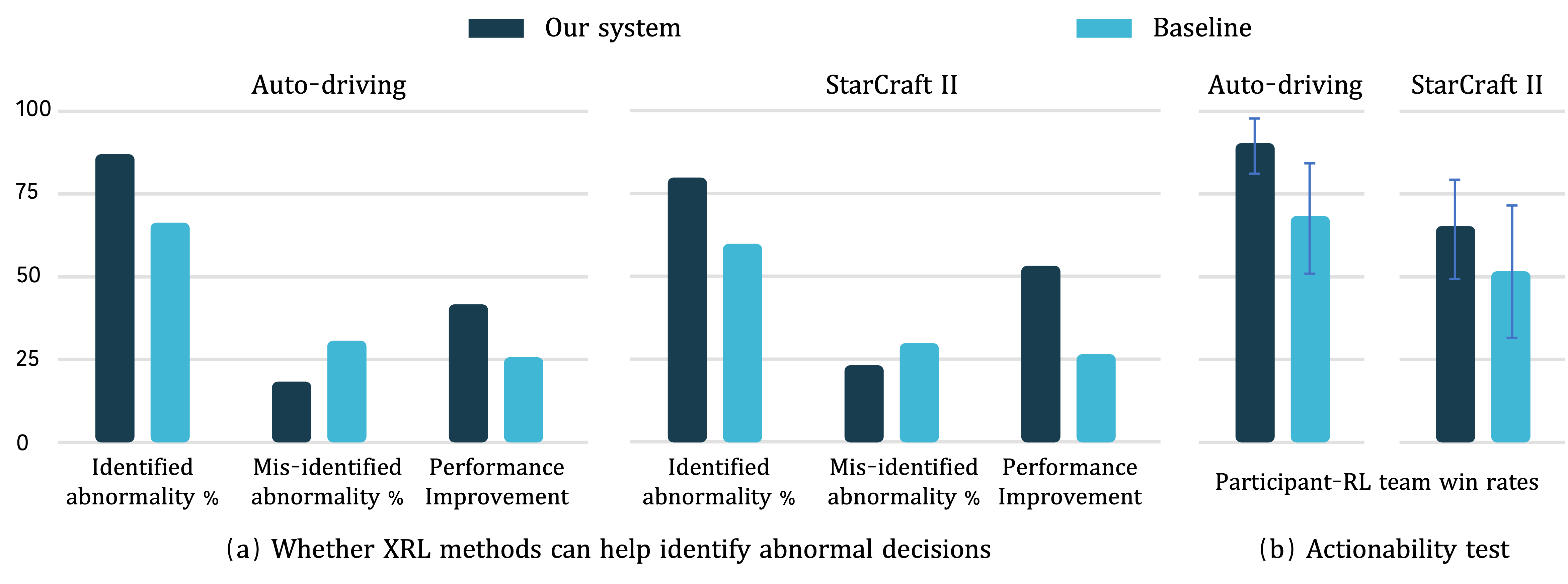}
    \caption{Left and Middle: Evaluating whether \name~can help RL developers achieve the first goal of using XRL (detect abnormal behavior of RL models). Right: the results of actionability test.}
    \label{fig:quantitative}
\end{figure}

\subsubsection{During Task.} 
To evaluate whether our system can help RL developers achieve the first goal, \emph{Abnormality detection}, we selected 10 time steps in each task and intentionally overrode the decisions of the RL agents at these steps. Participants were tasked to report the abnormal behaviors of RL agents after using the baseline and our system. The reported unexpected behaviors may not be those set by us, and we used a peer review method to decide whether these behaviors are indeed abnormal. We then reported (1) identified abnormality rate: the percentage of abnormal decisions that were identified, (2) misidentified abnormality rate: the percentage of reported unexpected decisions that were not abnormal, and (3) the change in RL agents' performance after changing the decisions reported by the participants.

To test whether our system can help the second goal, \emph{Limitation assessment}, developers are requested to answer a Likert question on a 7-point scale: "I find the XRL methods help me pinpoint some limitations of RL agents."

As for the third goal, \emph{End-user support}, after using the baseline and our system (without any intentional decision override), the end users were asked a Likert question on a 7-point scale: "I know in which situations the RL agent would work or not work, and know how to collaborate with it."

% The reported unexpected behaviors may not be those set by us, and we used a peer review method to decide whether these behaviors are indeed abnormal. We then reported (1) identified abnormality rate: the percentage of abnormal decisions that were identified, (2) misidentified abnormality rate: the percentage of reported unexpected decisions that were not abnormal, and (3) the change in RL agents' performance after changing the decisions reported by the participants. These two metrics evaluate whether an XRL method can fulfill \emph{Goal 1}. Participants in the first group then answered a Likert question on a 7-point scale: "I find the XRL methods help me pinpoint some limitations of RL agents." This question is for \emph{Goal 3}.

To evaluate whether our system helps achieve the goals from another perspective, we then carried out an \emph{actionability test} for all participants. In this test, we asked participants to replace an RL agent. Specifically, for auto-driving, participants controlled one of the vehicles while the others are still controlled by RL agents. Participants drove the vehicle for 10 episodes, each with a random starting point and a destination. We count the number of episodes where participants can reach the destination. For StarCraft II micromanagement, participants controlled one of ally units and collaborated with other RL agents to play 10 episodes. We report the win rate of this human-RL team. In these experiments, participants controlled the unit using keyboards. Intuitively, if the participants were more aware of the abnormal behaviors (\emph{Goal 1}) and limitations (\emph{Goal 2}) of RL models or trusted them appropriately (\emph{Goal 3}) after using an XRL tool, the win rates would be higher. 

\subsubsection{Post-Test.} 
% For the second group of participants, after using the baseline and our system (without any intentional decision override), they were asked a Likert question on a 7-point scale: "I know in which situations the RL agent would work or not work, and know how to collaborate with it." This question is to test whether our method can help achieve \emph{Goal 2}.

In order to evaluate whether our system can benefit to avoid the four techno-centric pitfalls (RQ2), all participants were asked four 7-point scale Likert questions: (1. For simplism) I think the explanation model gives sufficient details about RL agents' reasoning process. (2. For subjectivism) I find the explanations consistent with agents' behavior. (3. For formalism) I think the explanations are easy to understand. (4. For parochialism) I find the XRL model generates reasonable explanations for most situations. For RQ3, participants were asked a 7-point scale Likert question: (1) I like the interface and interactions provided by the XRL system. Then we took a ten-minute interview with each participant and talked about how they think about the component in our system and, if any, suggestions for improving the \name~system.

\section{Results}

\subsection{Quantitative Analysis}

\emph{Abnormality detection.} Results about whether XRL methods can help RL developers identify abnormal decisions are summarized in Fig.~\ref{fig:quantitative}. When using our system to explain auto-driving RL agents, 7 developers reported 94 unexpected decisions. Among the unexpected decisions, 61 are among the problematic decisions modified by us, i.e., $87.1\%$ of the abnormal decisions are detected. For the other $33$ unexpected decisions, $27$ of them are classified as abnormal by other RL practitioners, leading to a mis-identified abnormality rate of $18.2\%$. Moreover, after changing the reported unexpected decisions, the win rate of RL agents improved from $20.6\%$ to $62.3\%$. In comparison, the identified abnormality rate and mis-identified abnormality rate of the baseline XRL method on the auto-driving task is $66.2\%$ and $30.7\%$, respectively. Changing the reported unexpected decisions only improve the performance by $25.7\%$. Similar results can also be observed in the StarCraft II micromanagement task. These results demonstrate that our system can help RL developers better detect mistakes of RL agents. 

\begin{figure}
    \centering
    \includegraphics[width=\linewidth]{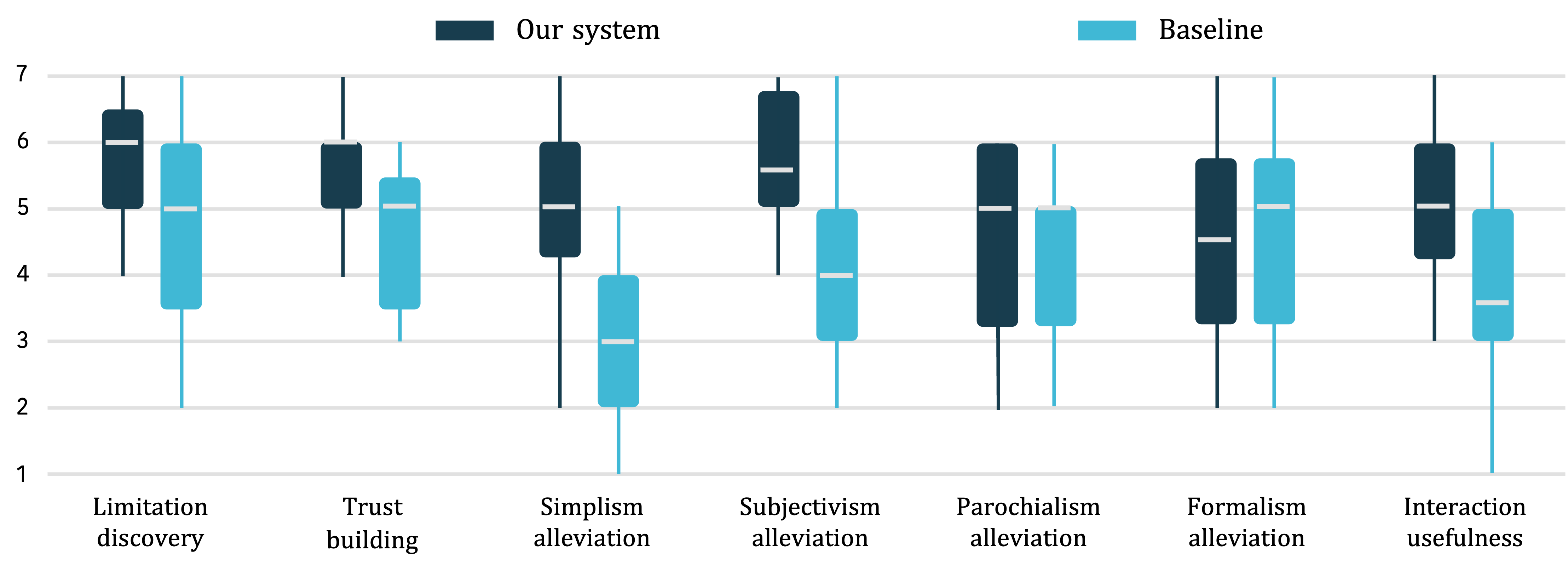}
    \caption{Evaluating whether our system can (1) help developers achieve the second (\emph{Limitation assessment}) and third (\emph{End-user support}) goals of using XRL; (2) help to avoid the four techno-centric pitfalls; and (3) provide useful interactions to system users.}
    \label{fig:qualitative}
\end{figure}

\emph{Actionability test.} In the task of auto-driving, after using our system, the participants can reach destinations without collisions in 90.3\% episodes. By contrast, the baseline can only help the users succeed in 68.4\% episodes. The StarCraft II micromanagement task is more challenging. After interacting with \name, participant-RL teams win 65.2\% of the games, while the baseline only achieves 51.5\%.

\emph{Limitation Assessment.} From the answers to the Likert question, users felt they could better pinpoint RL agents' limitations using our system (Mdn = 6.0, IQR = 5.0-6.5) than using the baseline (Mdn = 5.0, IQR = 3.5-6.0). 

\emph{End-user support.} End users also reported they know the success and failure cases of RL agents more clearly with our system (Mdn = 6.0, IQR = 5.0-6.0) than with the baseline (Mdn = 5.0, IQR = 3.5-6.0). 

\emph{Avoiding techno-centric pitfalls and system usefulness.} As for simplism, participants generally agreed that our system (Mdn = 5.0, IQR = 4.25-6.0) provides more details of the reasoning process of RL agents than the baseline (Mdn = 3.0, IQR = 2.0-4.0). As for subjectivism, participants reached a consensus that our system (Mdn = 5.5, IQR = 5.0-6.75) is less subjective than the baseline (Mdn = 4.0, IQR = 3.0-5.0). In terms of parochialism, the median performance of our system is the same as that of the baseline (Mdn = 5), but our system has a higher Q3 value (6.0 vs 5.0). The baseline algorithm (Mdn = 5.0, IQR = 3.25-5.75) performs slightly better than our system (Mdn = 4.5, IQR = 3.25-5.75) in terms of alleviating formalism. This is because the baseline only generates explanations but our system has more components for highlighting important history information. These results are summarized in Fig.~\ref{fig:qualitative}. However, these additional interactions are generally favored by the users, and they think the system is overall more more useful (Mdn = 5.0, IQR = 4.25-6.0) compared to the baseline (Mdn = 3.5, IQR = 3.0-5.0).

\subsection{Qualitative Analysis}

\subsubsection{\name~helps detect abnormality}

Participants' comments showed how our system achieves this goal. $P3$ said explanations for normal decisions are logically reasonable and mentioned the important role of highlighting influential observation information: "\emph{In my experiment on StarCraft II, I could easily distinguish abnormal decisions because the explanations for normal decisions are very logic but not for abnormal decisions. For example, when the health point is low, an RL agents tend to hide from enemies. If the system highlights a low health point but the action is attack, then something must went wrong there.}"

\subsubsection{\name~helps find limitations of RL models and build appropriate trust with end users.}

$P1$ found limitations of auto-driving agents in the online quick overview mode: "\emph{I got the sense that the agent turns the steering wheel more suddenly when it encounters a pedestrian (than the case of another vehicle). This may because the costs of colliding with pedestrians are more expensive. However, a gentle break is enough in most cases. Turning the steering wheel risks colliding with other objects.}" $P1$ also found the baseline didn't help him detect this issue, because "\emph{the causes of decisions are sometimes unclear and even absent.}" $P7$ found the StarCraft II agents sometimes lose due to a short-term attention, and she suggested changing the network architecture to deal with this problem: "\emph{\name~is like a temporal and spatial attention mechanism and one can gain important information for decisions. For example, I found StarCraft agents rarely care about information before three time steps ago, which results in missing some important information. I guess the underlying RL agents are using LSTM. Changing to GRU may solve this problem.}"

$P11$ thought our system is more transparent: "\emph{I like the way in which information information is labeled. And from watching replay, I knew that the RL agent would not brake when the speed is lower than 3 miles per hour unless the light turns red. This information was absent from the baseline. So I feel like I am more confident to interact after using \name.}" $P8$ mentioned that showing the possible results of a decision helped her a lot: "\emph{I was not familiar with StarCraft II, but seeing the results, for example, a unit is protected because another agent attracts more fire, made me understand the underlying decision-making process. If I know the results, I tend to trust RL.}" $P15$ found that "\emph{explanations showed that the RL model would always ignore a bike coming from the right side, and a collision would happen. I will be very careful when I use this model.}" Here, $P15$ discovered some failure cases, and he would not overly rely on the RL decisions, which is beneficial to build an appropriate trust.

\subsubsection{\name~better avoids the four techno-centric pitfalls.}

Many participants think our system can better avoid simplism. As $P10$ stated: "\emph{I believe that \name~provides the critical components that form a path leading the causes to the results. These will absolutely help me understand the reasoning process of RL agents.}" $P7$, who has a rich experience in the StarCraft II micromanagement tasks, also found our explanations reflect the motivation of decisions: "\emph{In this task, an important winning strategy is to focus fire, which means several allies attack an enemy simultaneously. I was impressed that \name~gives explanations mentioning this strategy. It truthfully reflects the real decision-making process of RL agents.}"

As for subjectivism, $11$ participants reported that they observed behaviors that were inconsistent with explanations generated by the baseline. For example, in $P3$'s case: "\emph{The baseline explained that the vehicle braked at an intersection because there was pedestrian walking along the road. However, according to my observation, the vehicle would brake slightly whenever it saw a traffic light. What's more, the pedestrian was walking along the road, I don't think the vehicle should consider it.}" These results are in line with our discussion in previous sections that using human-provided explanations as ground truth for training leads to subjectivism.

In terms of parochialism, we observed quantitatively that the median performance is the same for our system and the baseline, but our system has a higher Q3 value. We find this is because the baseline cannot explain Deep Q-learning agents' behavior well. Four participants were assigned a deep Q-learning agent, and all of them found the explanations of the baseline unsatisfactory. For example, $P14$ said: "\emph{I was aware of several obvious mistake even though I do not know anything about reinforcement learning. For example, the baseline says that the vehicle accelerates because the light is green while actually the light was red.}"

Although our method got a slightly lower rating for formalism alleviation, many RL practitioners agreed that explanations in natural language are much more understandable than explanations in other formats. For example, $P9$ said: "\emph{I used an XRL method based on finite-state automata in my work. Every time I give a presentation, audiences struggle to understand the explanations. I will use \name from now on.}"

\section{Discussion}

\subsection{Visualization RL training debugging}

Currently, our system is designed for generating explanations for trained RL models, but some participants of our user study mentioned in the interview that our system has the potential to help with RL model training, debugging, and fine-tuning. After detecting a decision abnormality, the next step for RL developers is to adjust their model or hyper-parameters for correcting the unexpected decisions. For instance, $P2$ suggested that we could provide a tool where users can freely change the network structure, fine-tune hyper-parameters, and visualize the resulting learning process. He held that RL development work would be much easier with such a tool.

\subsection{Building layered exploration structure}

One of the main novelties of \name~is allowing users to edit the observation information to check how decisions changed based on different inputs. However, we found that not all participants know how to use this active interaction function to change the parameters effectively. In addition, when the episode is long, users often do not know which parameters to start with, and such confusion affects users' rating of the usefulness of our \name. The possible reason is that the complexity of RL itself determines that many observation elements influence decision-making, and it is challenging to explore and manage a large number of elements.

To address this problem, we could build a layered exploration structure. We can design a tree-structured authoring interface to balance between ease-of-use and expressiveness. We can first show the default influential impact factors within the  surface-level UI. Then, every time when the user adjusts parameters to probe different aspects of the RL model, a new branch would be added to the UI. These selections and edits could be organized by a tree, through which the user can better track the components that have been explored.

\subsection{Exploring multiple factors for decisions}

During the user study, some participants mentioned that many factors could affect decisions. When generating the explanations, we only show one factor which exerts the most significant influence on the current decision. Nevertheless, there is no guarantee that other reasons will not matter. For example, sometimes, two factors have essentially the same impact on a decision. The support of multiple factor analysis could be achieved by setting a threshold to find all paths whose total weight is larger than this threshold, rather than running the longest path algorithm to find the causes with the most significant influence. In this way, we can flexibly change the explanation based on users' needs. From the perspective of UI design, we should add one more parameter setting box to allow users to adjust it easily so that our system supports more personalized demand on factor exploration.

\subsection{Limitations}

We identified two limitations of this \name~study. First, in the user study, we consider two tasks, auto-driving and StarCraft II micromanagement, which did not cover other applications. Meanwhile, the user study was carried out on simulators, and we did not evaluate RL developers' reactions to our system in real-world cases. Second, we did not evaluate the long-term experience of using our \name, and the study results may have been influenced by users’ initial impressions of the interface.

\section{Conclusion}
In this paper, we carry out a user-centric study for explainable reinforcement learning (XRL). XRL is important for RL developers who rely on XRL tools to better comprehend and work with opaque deep RL models. Prior XRL works, however, took a techno-centric approach to their research, neglecting whether they are useful for RL developers in practice. We identify four issues that widen the gap between current XRL approaches and the goals of practitioners to employ XRL methods. To solve these problems, we propose a novel explanation generation method based on counterfactual inference and build an XRL system surrounding this method. Our user study demonstrated that \name~could help RL developers better avoid the four pitfalls and achieve their goals for using XRL. For future work, it is a promising research direction to extend our work to the debugging and visualization of the learning process of reinforcement learning.

% In this paper, we study how to improve the performance of spoken dialogue interfaces. We decompose the uncertainty of spoken language into an entropy term measuring the structural uncertainty and a mutual information term reflecting the functional utility of the message. By optimizing these objectives within a reinforcement learning framework, for unseen tasks, our method can quickly learn task-specific communication protocols in natural language, which can help agents and human experimenters finish complex, temporally extended tasks.

% A drawback of our method is that the learned language is not perfectly precise. For example, it can not distinguish the shape of the object in the referential game. Moreover, on the 3D navigation task, communication among agents is noisy, occasionally sending messages describing connectivity of the house map or details of visual observation. We plan to look into these issues in the future.

\bibliographystyle{ACM-Reference-Format}
\bibliography{sample-base}

\appendix

\end{document}